%% file: Big.data.tex
\documentclass[conference]{IEEEtran}
\IEEEoverridecommandlockouts
\usepackage{cite}
\usepackage{amsmath,amssymb,amsfonts}
\usepackage{algorithmic}
\usepackage{graphicx}
\usepackage{textcomp}
\usepackage{xcolor}
\usepackage{tabularx} 
\usepackage{float}
\usepackage{hyperref}

\begin{document}

\title{JobPulse: A Big Data Approach to Real-Time Engineering Workforce Analysis and National Industrial Policy\thanks{This system was partially supported by the Microelectronics Commons Program, a DoD initiative, under award number N00164-23-9-G056.}}

% If needed, the subtitle note can be inserted like this:
% \author{\IEEEauthorblockN{\large{\textit{Note: Sub-titles are not captured in Xplore and should not be used}}}}

\author{\IEEEauthorblockN{Karen S. Markel}
\IEEEauthorblockA{\textit{College of Business} \\
\textit{University of Colorado Colorado Springs} \\
Colorado Springs, USA \\
kmarkel@uccs.edu}
\and
\IEEEauthorblockN{Mihir Tale}
\IEEEauthorblockA{\textit{Viterbi School of Engineering} \\
\textit{University of Southern California} \\
Los Angeles, CA, USA \\
tale@usc.edu}
\and
\IEEEauthorblockN{Andrea Belz}
\IEEEauthorblockA{\textit{Viterbi School of Engineering} \\
\textit{University of Southern California} \\
Los Angeles, CA, USA \\
abelz@usc.edu}
}

\maketitle

\begin{abstract}
Employment on a societal scale contributes heavily to national and global affairs; consequently, job openings and unemployment estimates provide important information to financial markets and governments alike. However, such reports often describe only the supply (employee job seeker) side of the job market, and skill mismatches are poorly understood. Job postings aggregated on recruiting platforms illuminate marketplace demand, but to date have primarily focused on candidate skills described in their personal profiles.  In this paper, we report on a big data approach to estimating job market mismatches by focusing on demand, as represented in publicly available job postings.  We use commercially available web scraping tools and a new data processing scheme to build a job posting data set for the semiconductor industry, a strategically critical sector of the United States economy; we focus on Southern California as a central hub of advanced technologies.  We report on the employer base and relative needs of various job functions.  Our work contributes on three fronts: First, we provide nearly real-time insight into workforce demand;  second, we discuss disambiguation and semantic challenges in analysis of employer data bases at scale; and third, we report on the Southern California semiconductor engineering ecosystem.
\end{abstract}

\begin{IEEEkeywords}
labor market intelligence, semiconductor industry, defense, Southern California, disambiguation 
\end{IEEEkeywords}

\section{Introduction}
National industrial policy typically serves multiple purposes, such as ensuring strategic capabilities and generating employment opportunities to citizens and residents. The first objective is naturally difficult to measure.  Investment in early-stage technology, particularly in the private sector, serves as a model by following outcomes such as early-stage technology portfolio evaluations \cite{Belz2021}, patent production \cite{Giga2021, Howell2017, Myers2022}, military procurement contracts \cite{Bhattacharya2021, Howell2021}, or subsequent private investment \cite{Howell2017, Lerner1999}.

The second objective – generating employment opportunities - is conceptually easier to track.  The Department of Labor (DoL) publishes regular employment reports, informing both the financial markets and public policy.  However, such reports only describe postings that are successfully filled, not the unfilled jobs; or unemployment reports provide information on job seekers.  Supply-side descriptive statistics of unfilled jobs can illuminate the labor market but systematic analysis is difficult to generate on a time scale relevant to financial markets and policymakers alike.  Ideally, information would go beyond the description such that employers and policy makers can identify tools to upskill the existing workforce, such as providing competencies, experience, academic or professional credentials based on industry needs; and motivate interest in employment opportunities in strategically important industries.

Boselli and collaborators \cite{Boselli2018} introduce the field of Labor Market Intelligence (LMI), which uses Artificial Intelligence and Large Language Models to inform labor market policy analysts, human resource managers, and workforce development specialists. The components of LMI include the systematic collection, processing, and analysis of various data sources such as job vacancies, industry trends and other relevant indicators \cite{Rahhal2024}, \cite{Tzimas2024}.

Timely estimates of unmet employment needs sensitive to job function, type of position, and job qualification (e.g. minimum requirements and preferred capabilities) are difficult to generate.  Often these efforts are implemented manually with benchmarking surveys geared toward hiring managers.  Such surveys ask questions about the organization's hiring needs, forecasted growth, and anticipated position qualifications.  Unfortunately, these studies can suffer from the challenges endemic to survey methodologies, such as low response rates and opportunistic sampling schemes that generate biased results.  

It is desirable to generate LMI by automating the collection of measurable job position needs for defined occupations.  In this paper, we approach workforce development needs (specifically unfilled positions) as a big data problem, and describe JobPulse, a prototype system that can report regionally unfilled job estimates in nearly real-time. As our use case, we consider the semiconductor industry in Southern California, a region of the United States recognized for its importance in developing technologies relevant to the aerospace and defense (A\&D) sectors \cite{Scott1993} and representing an important contribution to the national engineering landscape. 

In this work, we adopt the approach of Boselli and collaborators \cite{Boselli2018}.  We generate a data set by scraping publicly available job postings on a popular website targeted toward strategic recruiting and job seeker connections.  We describe the data extraction and processing to develop a dataset of job postings.

Our paper proceeds as follows. First, we briefly review research on the semiconductor labor market, data of interest, and traditional data collection techniques.  We then discuss the development of a data set through web scraping and data pre-processing.  Ultimately, we report descriptive statistics for a set of approximately 4,000 postings from 1,100 employers in Southern California.  This process illuminated disambiguation challenges previously unrecognized in the literature.   

We provide three important contributions to  engineering management scholarship.  For the policy community, we demonstrate and characterize a big data approach to an important question generally addressed with manual surveys or reported job placements.  Second, we add to the natural language processing literature, and specifically to bibliographic disambiguation, by reporting a previously unrecognized instance of deterministic, non-random effects that impact disambiguation processes.  Finally, we contribute to the field of management with a description of the links between two nationally strategic engineering fields – the semiconductor and defense industries - in Southern California, and we discuss the significance on a national scale.  Together, these contributions add broadly to semantic computing, engineering management, and industrial policy studies. 

\section{Workforce needs in the semiconductor industry}
The semiconductor industry workforce needs have been the subject of great interest in the United States.  In the period of 2019-2021, the COVID-19 pandemic prompted significant manufacturing and supply chain delays in computer chips, which cascaded through many industries, such as manufacturing of autos and home appliances.  Subsequently, the US Congress passed the CHIPS and Science Act (CHIPS Act) of 2022 – a major legislative policy - to invest in domestic semiconductor manufacturing. 

In the US, the Semiconductor Industry Association forecasts that nearly 4 million additional jobs requiring proficiency in technical fields will be created by 2030 \cite{SIAreport}.  Of those, an estimated one-third of the roles, such as skilled technicians, engineering, and computer science, are projected to go unfilled. Consequently, semiconductor workforce development (WFD) is an important aspect of programs funded by the CHIPS and Science Act.  One CHIPS Act WFD thread focuses on jobs related to the construction of manufacturing facilities.  The second thread, more critical to the industry’s outlook, is the expansion of a workforce able to develop, scale, and leverage the new technologies anticipated as the core capabilities of these manufacturing facilities. 

In 2023, the US Department of Defense created eight regional collaborations to spark growth in the domestic semiconductor industry, and as a long-standing anchor of the A\&D industry \cite{Scott1993}, Southern California is a natural home to one of these collaborations.  We can assess the general scale of employment in the semiconductor industry in this region.  The Federal Reserve Bank of St. Louis estimates that about 82,000 people were employed in the semiconductor industry in California April 2025 \cite{Fedreport}.  Industry reports suggest that about 20,000 of those  \cite{IBIS-Circuit, IBIS-Component, IBIS-Machine} are employed in the region of interest examined here.

\section{Online job postings to generate Labor Market Intelligence} 
Employee recruiting is designed to maximize a qualified applicant pool. Many online networking systems link recruiters and candidates broadly.  Professional networking applications have become a central tool for employers to distribute their job opportunities to potential applicants and establish an employer brand. 

Scholars have followed several approaches to scraping online job postings as a labor market proxy to classify occupational demand  \cite{Boselli2017, Tzimas2024}.  Boselli and collaborators \cite{Boselli2018} approach the use of job posting data from a macro-perspective, using a classification system of occupations to map to the job posting in a geographic region.  Templin and Hirsch found a strong correlation between online job postings and lagged hiring data, including identifying a discontinuity associated with the Great Recession of 2009 \cite{Templin2013}.  Job postings also align well with published DoL data \cite{GeorgetownReport}.  

\section{Data extraction}
The two fundamental steps required to develop a data set of job postings are searching through a corpus of potential postings, or crawling; and extracting the relevant information, or scraping.  The postings must satisfy some criteria (indexing terms) to be included in the potential data set.  JobPulse uses commercially available tools to match our indexing terms with specific fields in the job posting.  The indexing terms require a starting point. 

\subsection{Terminology and hierarchy}
We follow the classification scheme common in labor market research and define three levels at which job positions may be organized: 

\begin{itemize}
    \item A \textit{job title} describes a specific role within an organization. 
    \item A \textit{job family} captures similar job titles according to skills required, responsibilities, pay grade, and seniority. 
    \item A \textit{job function} aggregates job families related generally by education level and role in the organization's value chain. 
\end{itemize}

\subsection{Construction of the Set of Job-Specific Terms}

The Department of Defense collates a list of 31 job families relevant to the semiconductor industry, each of which comprises a set of 7-10 job titles. These terms are derived from the public O*NET database, used previously for LMI research \cite{Karakatsanis2017}.  For this analysis, we aggregated the 31 job families into four job functions: Scientists, Engineers, Technicians, and Organizational Support.

The combination of the job family names and the job title comprise the set of job-specific terms (JSTs).  Each JST is independently considered to be a keyword with which to run a search\footnote{A complete list of job titles for all job families is available upon request from the authors.}.  A total of 208 unique JSTs (31 job families and 177 job titles) were used. 

Importantly, the job posting platform is optimized for job seeker engagement.  Therefore, search platforms conduct a combination of term matching, personalized retrievals and semantic matching through a proprietary algorithm not observed by the platform user. The personalized retrieval output considers factors such as skills, the user's experience and education, search history, and the user's recent job applications and searches.  Thus, search terms are not considered in a strictly Boolean fashion. 

Initially, we ran two types of general searches: (1) JSTs alone, and (2) JSTs with the term ``semiconductor'' attached.  The results of these searches varied, depending on the nature of the JST.  For instance, a search on the term \textit{maintenance technician} generated many postings outside the semiconductor industry, whereas a search using the term \textit{layout engineer} was less likely to be confounded.  We viewed the results of these searches as a smaller universe of potential job postings.  This initial set of searches generated about 13,000 postings (Table \ref{tab:data_funnel}). 

Importantly, the JST methodology enabled us to find job postings even if the job title did not contain the JST itself. In other words, we searched for specific terms within the job posting, but not necessarily within the title. This broadened the search to find relevant postings that may have been overlooked with a strict job title search.  

\subsection{Data extraction tool} 
We used a commercially available tool to scrape a popular job posting site.   The tool included geographic filters, and we defined the Southern California region as consisting of one of three options:  Los Angeles metropolitan area (LA), Santa Barbara County (SB), or San Diego metropolitan area (SD).   We validated that this filter was operational by manually spot-checking many of the postings and confirming that they were listed in the areas of interest.  We also confirmed that the tool and the manual search generated the same results. Searches were conducted between March 15 and June 4, 2025.  Because the tool offered limited capability for chaining Boolean conditions, a single search was conducted with one "Search Phrase" (SP) in one region.  In this fashion, we constructed a database of all the SP in every region.  

\section{Data processing}
The output of the data extraction process consisted of roughly 13,000 observations (Table \ref{tab:data_funnel}).  Several processing steps followed: 

\input{tables/data_funnel.table}

\subsection{Removal of jobs in other industries}
We originally started with the JSTs considered as individual tokens rather than as n-grams of arbitrary length (typically two or three tokens long).  This enabled us to discover new titles previously unrecognized as relevant to the field. 

Later we defined a ``Search Phrase'' (SP) as follows:  the focal JST was embedded in quotation marks to ensure that it would be indexed as an n-gram and the word ``semiconductor'' was added to limit the search.  To meet the user engagement objectives described above, the platform search utilizes proprietary fuzzy match capabilities for single tokens, so that ``semiconductor'' or a fuzzy analogue might not have been in the job description but elsewhere in the company description.  This met our goal of finding specific JSTs in the broadly defined semiconductor industry.  For instance, the job family ``Product Engineer'' was considered as a JST and was coupled with the word 
``semiconductor'', such that the final search phrase was \textit{semiconductor ``product engineer''}. We constructed search phrases for each combination of ``semiconductor'' and a JST.  We validated that the JST appeared in the job description, even if not in the job title. 

We removed observations lacking the word ``semiconductor'' in the company and job descriptions.  This reduced our data set by about one-third, suggesting that the construction of an n-gram representing the JST as part of the Search Phrase was important to identify relevant observations. 

\subsection{JSTs at multiple hierarchy levels}
Some JSTs (e.g., \textit{Semiconductor packaging engineer}) were used both as a job family and as a job title.  In that case, the job family took precedence and the observation was assigned to that title within that family.  A JST used as a job family was removed from consideration as a job title in other job families.  

\subsection{Duplicated job identification numbers}
Each observation was labeled with a unique job identification number (job ID) generated by the platform.  However, several postings appeared multiple times in our data set, as the job description may have simultaneously contained several job titles.  For instance, a job description could contain both ``design engineer'' and ``layout engineer'' to describe the same role.  We sought to retain the equivalent of only one instance of each job ID, reasoning that it would be filled by a single person.  

When a jobID contained multiple JSTs, it was weighted equally for each JST.  For instance, in the above example, the job description would be weighted with 0.5 as a ``design engineer'' and 0.5 as a ``layout engineer.''  This process reduced our data set by more than half.  We review the implications of this phenomenon in our Discussion\footnote{An implication with minimal impact was that the total employment needs tables may be subject to rounding errors.}.

\subsection{Multiple job identification numbers}
In some cases, a job posting was listed in multiple locations, and each was assigned a different job ID.  When this occurred, each observation was treated as a distinct job posting and counted independently in each region where it appeared.  This approach ensures that our dataset reflects the actual number of open positions across all geographic regions.

\subsection{Discovery of relevant job titles}
In a small number of cases, the search process found titles that were not in our list of JSTs.  For instance, review of the ``RF Engineer'' description revealed that it was a ``radio frequency engineer'' with a high degree of overlap to a ``design engineer''.  These functions are not equivalent but are certainly related  For this reason, several titles, ``RF engineer'' and ``radar engineer'' were added to the ``design engineer'' job family.

\section{Results}

\subsection{Overview of job postings}
As noted earlier, we aggregated the 31 job families into four broad job functions: Scientist, Engineer,  Technician, and  Operational Support.  We report the distribution of job titles within the job functions and find that roughly 4,000 job postings (Table \ref{tab:jobs_summary_table}) were available in the Southern California semiconductor industry in the spring of 2025.  Approximately three-fourths were in Los Angeles, with the remainder split among Santa Barbara and San Diego.  Importantly, the summary numbers already indicate that prior estimates \cite{SIAreport} may not accurately reflect the needs in Southern California as of 2025.  Specifically, we find that the relative need for technicians and engineers is about 1:3, in stark contrast to the 2:1 numbers previously reported. 

Another important stylized fact is that this summary can be compared with the current estimated regional workforce of 20,000 employees described in the Introduction.  The need for 4,000 employees suggests a regional shortfall of about 20\%.
\input{tables/summary.table}

\subsection{Scientists}
The Scientist function comprises three job titles (Table \ref{tab:scientist_keyword_summary}).  Of those, approximately three-fourths are computer scientists, demonstrating the ongoing importance of data science throughout the scientific enterprise.  While ``Materials Research Scientist'' may have a small number of postings, the requirements overlap heavily with the more general ``Research Scientist'' and may be captured therein.  Scientist job requirements typically include an advanced degree and extensive experience. 

\input{tables/scientist_keyword_summary.table}

\subsection{Engineers}
The greatest need is in the Engineer job function, comprising about two-thirds of all jobs in this field.  The distribution across job families is highly variable, with ``Product Engineer'', ``Design Engineer'', and ``Quality Assurance Engineer'' jointly representing one-third of the job vacancies (Table \ref{tab:engineer_keyword_summary}).

\input{tables/engineer_keyword_summary.table}

\subsection{Technicians}
Initially only one job family, Fab Technician, was assigned to this category.  However, we found that about 10\% of the relevant jobs did not use any of the titles in the DOD list and instead were discovered using ``Microelectronics Technician'', which added another 10\% to our estimated Technician total.  In sum, we report that Technicians represent about 20\% of the estimated job demand in the semiconductor industry in Los Angeles (Table \ref{tab:technician_keyword_summary}). 
\input{tables/technician_keyword_summary.table}

\subsection{Operational Support}
A final job function in the semiconductor industry is Operational Support.  Importantly, these job families (Supply Chain Analysts and System Administrator) are most likely to be used in other industries, and consequently are least refined.  The current estimate of families comprising postings in organizations that aligned with ``Semiconductor'' - is that these roles represent a small fraction (less than 5\%) of the need in the semiconductor industry (Table \ref{tab:operation_keyword_summary}. 

\input{tables/operation_keywords_summary.table}

\subsection{The employer base}
The nature of our data set naturally leads to general insights regarding the employer base.  The first is the number of companies represented in this sample.  At first glance, 1,269 companies offered the 4,044 positions listed here. Clearly, this far exceeds the reach of surveys and related manual processes.  However, this estimated employer base is confounded by two factors: 
\begin{itemize}
\item \textit{Corporate divisions.}  In some cases, we identified multiple divisions of the same company listing jobs - for instance, both Amazon and Amazon Web Services were listed as employers.  Approximately 15\% of the employers are divisions of a larger company.  
\item \textit{Frequently used terms in corporate names.} A greater challenge is the frequent use of specific words in company names - for instance, ``American'' and ``Advanced'' are used frequently.  This phenomenon, onomastic profusion, has been identified as an assignee disambiguation challenge in patents and other publications \cite{Belz2023-onomastic}.  It stems from a corporate branding strategy that exploits the semantic content of specific words to generate company names \cite{Motschenbacher2020}, such that specific words appear with high frequency in a database of names.  As this is a strategic choice, it forms a source of error that is not random - for instance, a study of small companies found that companies with names that included ``Research'' or ``Science'' were more likely to win federal small business research awards \cite{Belz2023-onomastic}.  Here we document for the first time an extension of onomastic profusion beyond bibliographic studies to LMI. 
\end{itemize}

An automated detection of such terms is needed in order to distinguish the companies easily.  When the data base is constructed, each company name is naturally a multigram of arbitrary length.  This list had certain properties that we could exploit to disambiguate company names as follows: 

\begin{itemize}
\item \textit{First word extraction.}  We extracted the first word of each company name.
\item \textit{Development of a name dictionary.}  Review of the list of employer names revealed that specific common words were used frequently as the first word. In some cases, this was true for organizational reasons (e.g., ``University of''); and in others, due to the onomastic profusion described above (e.g., ``American'' or ``Advanced.'')  We developed a dictionary to identify these cases.  
\item \textit{Next word extraction (multigrams).}  For multigrams, the next word was also extracted and compared.  This enabled the disambiguation of multigrams such as Advanced Micro and Advanced Systems.  This process was continued as necessary, such as to disambiguate University of California Los Angeles and University of California Santa Barbara.
\end{itemize}

This process reduced the number of unique employers by less than 10\% - from 1,269 to 1,135 employers.  Notably, we shed light on the semiconductor employer base in Southern California.  The top individual recruiters in Southern California were Anduril, SpaceX, and Relativity Space, comprising a total of 433 job postings. These are newer companies in the defense industrial base (DIB) and thus they may have a greater need than incumbents if they are augmenting a smaller baseline level of employees.  Regardless, this suggests that Southern California uses a semiconductor workforce to support the DIB.

On a related note, we see that the top individual recruiters comprise just over 10\% of our sample.  Evidently the semiconductor industry in Southern California is severely fragmented, with a mean number of job postings per employer of 3.6. We report the stylized finding that most companies are looking for very few people.  This could significantly impact the reach of potential policy interventions to provide a greater workforce through funded collaborations such as the Hubs, as it would be obviously difficult to reach the entire employer base.  Even a 10\% penetration would require managing a collaboration of 100 firms, well beyond the reach of a typical single federal award.

\section{Limitations and future research}
This work is in its infancy and clearly has several limitations.  A simple concern is that the commercial tool that we used could only access data on a single platform.  It would be useful to triangulate across platforms to see which jobs are distributed most broadly, as this would provide a measure of need.  Likewise, the platform to which we had access may be disproportionately used by job seekers of certain educational levels, thus generating a selection bias.  Our level of concern regarding this potential bias is mitigated by the fact that we saw all job functions represented.  The general qualitative trends were supported by industry interviews. 

In addition, our stringent filtering approach is sensitive to the size of the JST repository.  As Rahhal and collaborators \cite{Rahhal2024} note, the job title alone without the job description can result in incomplete information.  Our methodology searches beyond the title to the entire job description; however, our estimates still assume that the JST list is comprehensive.  Our discovery of the term ``Microelectronics Technician'' suggests that this assumption is minimally incorrect.  The magnitude of the incremental addition to the list of postings suggests that this is not a significant concern.

Another potential limitation is that while we restricted our search geographically, we did not filter the postings by work from home (WFH), hybrid, or other job location constraints.  It is feasible that a job is posted in San Diego but could be executed in New York.  We reason that this is more likely to be true for the Organizational Support functions, as they do not require the laboratory facilities of semiconductor fabrication.  

However, some Scientist roles - in particular, those associated with development of Electronic Design Automation (EDA) tools - could conceivably be conducted remotely.  One estimate of employment growth suggests that EDA tool employers will grow at nearly 7\% \cite{IBIS-EDA}, in contrast to the 0-4\% estimated for conventional semiconductor manufacturing.  As  EDA tools are more likely to require graduate scientific training, this could generate further stratification of employment opportunities by educational level and by the ability to WFH. 

To date we have not yet processed the job descriptions beyond simply retrieving the JST as a keyword in the text.  The job description is clearly rich with information regarding educational requirements, as well as other data.  

Additionally, our methodology revealed the duplication of job postings with multiple relevant job titles.  Conceivably an employer embeds more JSTs to create a many-to-one mapping - namely, the employer uses 
more JSTs to increase the possibility that a candidate finds the posting.  Thus it could be a measure of need. Further studies would be useful to link this with other observables, such as the vacancy duration \cite{Chen2023}. \cite{Tzimas2024} suggests that industry analyses can be challenged if job (titles) may be found in other industries.  For example, in our study, the analysis of technician job postings, needed to be broadened to include more relevant job posting data into our analysis.  Given the demand, skills and expertise needed for the semiconductor industry, it may be important to delve deeper into the competitive environment to recruit this talent outside those often included in national conversations (e.g. large defense contractors). 

% \FIX{Karen will cite Tzimas \cite{Tzimas2024} - in finishing this}

\section{Conclusion}
The development of LMI through automated construction of a job posting list has important implications for labor market estimations.  A robust job posting list enables analysis by characteristics of interest (e.g., industry, region, occupation, education requirements).  Thus such a database can illuminate various trends and may even anticipate broad labor market developments.  For this reason, linking the JST list to established job function studies is an important product of this research. 

We make three major contributions.  First, we demonstrate an automated method to reveal hiring trends in nearly real-time.  This information can help steer federal workforce monitoring and even investment appropriately.  Moreover, it is even feasible to develop a tool that could  inform financial markets as a supplement to reports published by federal agencies. 

Second, we report on a previously unrecognized disambiguation challenge by finding words that are more commonly used in company names. This adds to the rich semantic computing literature on bibliographic disambiguation, and more importantly forms an underappreciated source of potential bias.  

Finally, we describe the labor market gaps in Southern California, historically a leading center for aerospace and defense applications of the semiconductor industry \cite{Scott1993}.  We do not find support for previous estimates of need in the semiconductor industry \cite{SIAreport}.  Specifically, we find that the technician-to-engineer gaps in Southern California are of order 1:3, contrasting greatly with   the projected ratio 2:1 ratio. This has critical implications for WFD programs funded by the federal and local governments.   

\section*{Acknowledgment}
The authors thank Hayley Atwater, Steve Crago, Lida Dimitropoulou, Erin Gawron-Hyla, Larry Kushner, Zach Lemnios, Victor Mai, Charles Parker, and Kari Quan.

\bibliographystyle{IEEEtran}
\bibliography{Big.Data}

\vspace{12pt}
\color{red}
\end{document}

%% file: tables/data_funnel.table.tex
% latex table generated in R 4.5.0 by xtable 1.8-4 package
% Thu Jun  5 11:02:45 2025
\begin{table}[ht]
\centering
\caption{Data Filtering Process}
\begin{tabularx}{\columnwidth}{lrrrr}
  \hline
Data Funnel & LA & SB & SD & Total \\
  \hline
Observations & 10849 & 697 & 1443 & 12989 \\
  Removed jobs in other industries & 7597 & 599 & 1160 & 9356 \\
  Removed duplicates & 3057 & 301 & 688 & 4046 \\ 
   \hline
\end{tabularx}
\label{tab:data_funnel}
\end{table}

%% file: tables/summary.table.tex
% latex table generated in R 4.5.0 by xtable 1.8-4 package
% Thu Jun  5 11:29:55 2025
\begin{table}[ht]
\centering
\caption{Unmet Employment Needs by Job Function}
\begin{tabularx}{\columnwidth}{Xrrrr}
  \hline
Job Function & LA & SB & SD & Total \\ 
  \hline
    Scientist & 404 & 32 & 66 & 502 \\
    Engineer & 1973 & 204 & 416 & 2593 \\
    Technician & 561 & 59 & 163 & 783 \\
  Operational support & 118 & 5 & 43 & 166 \\
  \hline
  Total & 3056 & 300 & 688 & 4044 \\
   \hline
\end{tabularx}
\label{tab:jobs_summary_table}
\end{table}

%% file: tables/scientist_keyword_summary.table.tex
% latex table generated in R 4.5.0 by xtable 1.8-4 package
% Tue Jun 10 23:52:33 2025
\begin{table}[ht]
\centering
\caption{Unmet Needs for Scientists}
\begin{tabularx}{\columnwidth}{Xrrrr}
  \hline
Job Family & LA & SB & SD & Total \\
  \hline
Computer Scientist & 292 & 22 & 46 & 360 \\
  Research Scientist & 108 & 10 & 20 & 138 \\
  Materials Research Scientist & 3 & 0 & 0 & 3 \\
  \hline
   Total & 403 & 32 & 66 & 501 \\ 
   \hline
\end{tabularx}
\label{tab:scientist_keyword_summary}
\end{table}

%% file: tables/engineer_keyword_summary.table.tex
% latex table generated in R 4.5.0 by xtable 1.8-4 package
% Tue Jun 10 23:52:33 2025
\begin{table}[ht]
\centering
\caption{Unmet Needs for Engineers}
\resizebox{\columnwidth}{!}{%
\begin{tabular}{lrrrr}
  \hline
Job Family & LA & SB & SD & Total \\
  \hline
Product Engineer & 312 & 41 & 93 & 446 \\ 
  Quality Assurance Engineer & 232 & 12 & 29 & 273 \\
  Design Engineer & 204 & 25 & 44 & 273 \\
  Process Engineer & 132 & 19 & 26 & 177 \\
  Embedded Security Engineer & 121 & 20 & 25 & 166 \\
  Electronic Design Automation (EDA) Engineer & 89 & 13 & 30 & 132 \\
  PCB Design Engineer & 78 & 9 & 16 & 103 \\ 
  Test Engineer & 68 & 7 & 17 & 92 \\
  Embedded Systems Engineer & 63 & 6 & 19 & 88 \\
  Packaging Engineer & 65 & 2 & 18 & 85 \\
  Safety Engineer & 67 & 5 & 9 & 81 \\
  Environmental Engineer & 56 & 7 & 10 & 73 \\
  Validation Engineer & 51 & 2 & 13 & 66 \\ 
  Security Engineer & 46 & 7 & 10 & 63 \\
  Firmware Engineer & 46 & 3 & 13 & 62 \\
  Field Application Engineer & 48 & 7 & 5 & 60 \\
  Process Design Kit (PDK) Engineer & 48 & 4 & 5 & 57 \\
  Reliability Engineer & 47 & 2 & 6 & 55 \\
  Failure Analysis Engineer & 49 & 1 & 2 & 52 \\
  Physical Design Engineer & 40 & 3 & 8 & 51 \\ 
  Mask Engineer & 40 & 2 & 6 & 48 \\
  Photolithography Engineer & 37 & 1 & 5 & 43 \\
  Metrology Engineer & 29 & 1 & 6 & 36 \\
  Yield Enhancement Engineer & 5 & 3 & 1 & 9 \\
  \hline
   Total & 1973 & 202 & 416 & 2591 \\
   \hline
\end{tabular}
}
\label{tab:engineer_keyword_summary}
\end{table}

%% file: tables/technician_keyword_summary.table.tex
% latex table generated in R 4.5.0 by xtable 1.8-4 package
% Tue Jun 10 23:52:33 2025
\begin{table}[ht]
\centering
\caption{Unmet Needs for Technicians}
\begin{tabularx}{\columnwidth}{Xrrrr}
  \hline
    Job Family & LA & SB & SD & Total \\
  \hline
    Fab Technician & 480 & 59 & 157 & 696 \\
    Microelectronics Technician & 81 & 0 & 7 & 88 \\
  \hline
  Total & 561 & 59 & 164 & 784 \\
   \hline
\end{tabularx}
\label{tab:technician_keyword_summary}
\end{table}

%% file: tables/operation_keywords_summary.table.tex
% latex table generated in R 4.5.0 by xtable 1.8-4 package
% Tue Jun 10 23:52:33 2025
\begin{table}[ht]
\centering
\caption{Unmet Needs for Operational Support}
\begin{tabularx}{\columnwidth}{Xrrrr}
  \hline
Job Family & LA & SB & SD & Total \\ 
  \hline
Supply Chain Analyst & 76 & 2 & 35 & 113 \\
  Systems Administrator & 43 & 3 & 8 & 54 \\
  \hline
   Total & 119 & 5 & 43 & 167 \\
   \hline
\end{tabularx}
\label{tab:operation_keyword_summary}
\end{table}

%% file: Big.data.bbl
\begin{thebibliography}{10}
\providecommand{\url}[1]{#1}
\csname url@samestyle\endcsname
\providecommand{\newblock}{\relax}
\providecommand{\bibinfo}[2]{#2}
\providecommand{\BIBentrySTDinterwordspacing}{\spaceskip=0pt\relax}
\providecommand{\BIBentryALTinterwordstretchfactor}{4}
\providecommand{\BIBentryALTinterwordspacing}{\spaceskip=\fontdimen2\font plus
\BIBentryALTinterwordstretchfactor\fontdimen3\font minus
  \fontdimen4\font\relax}
\providecommand{\BIBforeignlanguage}[2]{{%
\expandafter\ifx\csname l@#1\endcsname\relax
\typeout{** WARNING: IEEEtran.bst: No hyphenation pattern has been}%
\typeout{** loaded for the language `#1'. Using the pattern for}%
\typeout{** the default language instead.}%
\else
\language=\csname l@#1\endcsname
\fi
#2}}
\providecommand{\BIBdecl}{\relax}
\BIBdecl

\bibitem{Belz2021}
\BIBentryALTinterwordspacing
A.~Belz, R.~J. Terrile, F.~Zapatero, M.~Kawas, and A.~Giga, ``Mapping the
  “valley of death”: Managing selection and technology advancement in
  nasa’s small business innovation research program,'' \emph{IEEE
  Transactions on Engineering Management}, vol.~68, no.~5, p. 1476–1485, Oct.
  2021. [Online]. Available: \url{http://dx.doi.org/10.1109/TEM.2019.2904441}
\BIBentrySTDinterwordspacing

\bibitem{Giga2021}
\BIBentryALTinterwordspacing
A.~Giga, A.~Graddy-Reed, A.~Belz, R.~J. Terrile, and F.~Zapatero, ``Helping the
  little guy: the impact of government awards on small technology firms,''
  \emph{The Journal of Technology Transfer}, vol.~47, no.~3, p. 846–871, May
  2021. [Online]. Available: \url{http://dx.doi.org/10.1007/s10961-021-09859-0}
\BIBentrySTDinterwordspacing

\bibitem{Howell2017}
\BIBentryALTinterwordspacing
S.~T. Howell, ``Financing innovation: Evidence from r\&d grants,''
  \emph{American Economic Review}, vol. 107, no.~4, p. 1136–1164, Apr. 2017.
  [Online]. Available: \url{http://dx.doi.org/10.1257/aer.20150808}
\BIBentrySTDinterwordspacing

\bibitem{Myers2022}
\BIBentryALTinterwordspacing
K.~R. Myers and L.~Lanahan, ``Estimating spillovers from publicly funded r\&d:
  Evidence from the us department of energy,'' \emph{American Economic Review},
  vol. 112, no.~7, p. 2393–2423, Jul. 2022. [Online]. Available:
  \url{http://dx.doi.org/10.1257/aer.20210678}
\BIBentrySTDinterwordspacing

\bibitem{Bhattacharya2021}
\BIBentryALTinterwordspacing
V.~Bhattacharya, ``An empirical model of r\&d procurement contests: An analysis
  of the dod sbir program,'' \emph{Econometrica}, vol.~89, no.~5, p.
  2189–2224, 2021. [Online]. Available:
  \url{http://dx.doi.org/10.3982/ECTA16581}
\BIBentrySTDinterwordspacing

\bibitem{Howell2021}
J.~V.~R. Sabrina T.~Howell, Jason~Rathje and J.~Wong, ``{Opening Up Military
  Innovation: Causal Effects of Reforms to US Defense Research},'' National
  Bureau of Economic Research, Tech. Rep., 2021.

\bibitem{Lerner1999}
\BIBentryALTinterwordspacing
J.~Lerner, ``The government as venture capitalist: The long‐run impact of the
  sbir program,'' \emph{The Journal of Business}, vol.~72, no.~3, p. 285–318,
  Jul. 1999. [Online]. Available: \url{http://dx.doi.org/10.1086/209616}
\BIBentrySTDinterwordspacing

\bibitem{Boselli2018}
R.~Boselli, M.~Cesarini, F.~Mercorio, and M.~Mezzanzanica, ``Classifying online
  job advertisements through machine learning.'' \emph{Future Generation
  Computer Systems}, vol.~86, pp. 319--328, 2018.

\bibitem{Rahhal2024}
I.~Rahhal, I.~Kassou, and M.~Ghogho, ``Data science for job market analysis: A
  survey on applications and techniques.'' \emph{Expert Systems with
  Applications}, 2024.

\bibitem{Tzimas2024}
\BIBentryALTinterwordspacing
G.~Tzimas, N.~Zotos, E.~Mourelatos, K.~C. Giotopoulos, and P.~Zervas, ``From
  data to insight: Transforming online job postings into labor-market
  intelligence,'' \emph{Information}, vol.~15, no.~8, p. 496, Aug. 2024.
  [Online]. Available: \url{http://dx.doi.org/10.3390/info15080496}
\BIBentrySTDinterwordspacing

\bibitem{Scott1993}
\BIBentryALTinterwordspacing
A.~J. Scott, ``Interregional subcontracting patterns in the aerospace industry:
  The southern california nexus,'' \emph{Economic Geography}, vol.~69, no.~2,
  p. 142, Apr. 1993. [Online]. Available:
  \url{http://dx.doi.org/10.2307/143533}
\BIBentrySTDinterwordspacing

\bibitem{SIAreport}
``State of the u.s. semiconductor industry,'' Semiconductor Industry
  Association, Tech. Rep., 2024,
  {\url{https://www.semiconductors.org/wp-content/uploads/2024/10/SIA\_2024\_State-of-Industry-Report.pdf}}.

\bibitem{Fedreport}
F.~R.~B. of~St.~Louis, ``All employees: Manufacturing: Durable goods:
  Semiconductor and other electronic component manufacturing in california,''
  2025, https://fred.stlouisfed.org/series/SMU06000003133440001SA, retrieved on
  June 13, 2025.

\bibitem{IBIS-Circuit}
IBISWorld, ``Semiconductor \& circuit manufacturing in california,'' 2025,
  report CA33441A.

\bibitem{IBIS-Component}
------, ``Circuit board \& electronic component manufacturing in california,''
  2025, report CA33441B.

\bibitem{IBIS-Machine}
------, ``Semiconductor machinery manufacturing in california,'' 2025, report
  CA33329A.

\bibitem{Boselli2017}
\BIBentryALTinterwordspacing
R.~Boselli, M.~Cesarini, S.~Marrara, F.~Mercorio, M.~Mezzanzanica, G.~Pasi, and
  M.~Viviani, ``Wolmis: a labor market intelligence system for classifying web
  job vacancies,'' \emph{Journal of Intelligent Information Systems}, vol.~51,
  no.~3, p. 477–502, Sep. 2017. [Online]. Available:
  \url{http://dx.doi.org/10.1007/s10844-017-0488-x}
\BIBentrySTDinterwordspacing

\bibitem{Templin2013}
T.~Templin and L.~Hirsch, ``Do online job ads predict hiring?'' New York City
  Labor Market Information Service, Tech. Rep., 2013,
  https://www.gc.cuny.edu/sites/default/files/2022-01/CUR-Do-Online-Ads-Predict-Hiring.pdf.

\bibitem{GeorgetownReport}
A.~P. Carnevale, T.~Jayasundera, and D.~Repnikov, ``Understanding online job
  ads data,'' Georgetown University Center on Education and the Workforce,
  Tech. Rep., 2014.

\bibitem{Karakatsanis2017}
\BIBentryALTinterwordspacing
I.~Karakatsanis, W.~AlKhader, F.~MacCrory, A.~Alibasic, M.~A. Omar, Z.~Aung,
  and W.~L. Woon, ``Data mining approach to monitoring the requirements of the
  job market: A case study,'' \emph{Information Systems}, vol.~65, p. 1–6,
  Apr. 2017. [Online]. Available:
  \url{http://dx.doi.org/10.1016/j.is.2016.10.009}
\BIBentrySTDinterwordspacing

\bibitem{Belz2023-onomastic}
\BIBentryALTinterwordspacing
A.~Belz, A.~Graddy-Reed, F.~Shweta, A.~Giga, and S.~M. Murali, ``Deterministic
  bibliometric disambiguation challenges in company names,'' in \emph{2023 IEEE
  17th International Conference on Semantic Computing (ICSC)}.\hskip 1em plus
  0.5em minus 0.4em\relax IEEE, Feb. 2023, p. 239–243. [Online]. Available:
  \url{http://dx.doi.org/10.1109/ICSC56153.2023.00047}
\BIBentrySTDinterwordspacing

\bibitem{Motschenbacher2020}
\BIBentryALTinterwordspacing
H.~Motschenbacher, ``Corpus linguistic onomastics: A plea for a corpus-based
  investigation of names,'' \emph{Names}, vol.~68, no.~2, p. 88–103, Apr.
  2020. [Online]. Available:
  \url{http://dx.doi.org/10.1080/00277738.2020.1731240}
\BIBentrySTDinterwordspacing

\bibitem{IBIS-EDA}
IBISWorld, ``Electronic design automation software developers in the us,''
  2025, report OD4540.

\bibitem{Chen2023}
\BIBentryALTinterwordspacing
C.-W. Chen and L.~Y. Li, ``Is hiring fast a good sign? the informativeness of
  job vacancy duration for future firm profitability,'' \emph{Review of
  Accounting Studies}, vol.~28, no.~3, p. 1316–1353, Aug. 2023. [Online].
  Available: \url{http://dx.doi.org/10.1007/s11142-023-09797-2}
\BIBentrySTDinterwordspacing

\end{thebibliography}
